\documentclass[preprint]{aastex}
\slugcomment{submitted to AJ}
\shorttitle{Stellar Populations in Disks.III...}
\shortauthors{Moll\´{a} et al.}
\input epsf
\usepackage{graphicx} 
\begin{document}

\title{The Stellar Populations of Spiral Disks. III. Constraining their 
evolutionary histories}

\author{Mercedes Moll\'{a}}
\affil{Departamento de F\'{\i}sica Te\'{o}rica, Universidad Aut\'{o}noma de
 Madrid, 28049 Cantoblanco, Spain}
\email{mercedes.molla@uam.es}

\author{Eduardo Hardy}
\affil{National Radio Astronomy Observatory 
\footnote{The National Radio Astronomy Observatory is a facility of
the National Science Foundation operated under cooperative agreement
by Associated Universities, Inc., U.S.A.}}

\email{ehardy@nrao.edu}

\received{}
\revised{}
\accepted{}

\begin{abstract}

 We study the old problem of the uniqueness of chemical evolution
models by analyzing a set of multiphase models for the galaxy NGC~4303
computed for a variety of plausible physical input parameters.

We showed in \citet{mol99} that multiphase chemical evolution models
for the three Virgo cluster galaxies NGC~4303, NGC~4321 and NGC~4535,
were able to reproduce the observed radial distributions of spectral
indices Mg$_{2}$ ad Fe52. Chemical evolution models may, however, fit
the present-epoch radial distributions with different star formation
histories, thus we need to include time-dependent constraints to the
problem. The two spectral indices above depend on the time--averaged
history of star formation, but are in turn affected by the well known
age-metallicity degeneracy which prevents the disentangling of age and
metallicity for stellar populations, another uniqueness problem we also 
discuss. 

Our aim is to determine if the possible input parameters may be
strongly constrained when both radial distributions (nebular and
stellar) are used. In order to accomplish this we run a large number
of models (500) for NGC~4303 varying the model input parameters. Less
than 4 \% of the models (19) fit the present day observational data
within a region of 95\% probability.  The number of models reduces to
a $\sim 1$ \% (6) when we also ask them to reproduce the time-averaged
abundances represented by the spectral indices.  Thus, by proving that
only a small fraction of the models are favored in reproducing the
present day radial abundance distributions and the spectral indices
data simultaneously, we show that these spectral indices provide
strong time-dependent additional constraints to the possible star
formation and chemical histories of spiral disks.
\end{abstract}

\keywords{ galaxies: abundances -- galaxies: evolution-- galaxies:
spiral --galaxies: stellar content}

\section{Introduction}

Variations of abundances with galactocentric radius have been widely
observed in external spiral galaxies \citep{dia89,ski96,gar97}, mostly
{\sl via} observations of H\,{\sc ii} regions which represent the
abundances in the gas phase at the present time. Since \citet{tin80},
a large number of chemical evolution models have been developed, by
including different physical mechanisms \citep[see][]{got92,kop94} in
order to reproduce these radial gradients.

In this work we use a multiphase chemical evolution model (MCEM) where
an infall dependent on galactocentric distance and a two-step star
formation law, whereby molecular clouds are formed before stars, are
the basic hypotheses. The outcome is star formation simulating a power
law in the gas density with an exponent $n>1$.  This model was first
applied to the Solar Neighborhood \citep{fer92}, to the whole Galactic
(MWG) disk \citep[][hereafter FMPD]{fer94}, and to the bulge
\citep{mol95}. Next the same model was applied to disks \citep{mol96},
and bulges \citep{mol00} of a set of spiral galaxies of different
morphological types, the observed radial distributions of diffuse and
molecular gas, oxygen abundances and star formation rate being
successfully reproduced. It seems therefore that the MCEM produces
results in good agreement with the data representing the present-day
interstellar medium.

It is well known, from studies of the MWG, however, that by tuning in
an appropriate way both the history of star formation and the infall
rate it is possible to reproduce the same present-time radial
distributions using very different evolutionary scenarios \citep[see,
for examples and further references,][]{mol92,tos96}. \citet{tos88}
early addressed this problem of the uniqueness of theoretical chemical
evolution models. In fact, in most cases differences in this kind of
models have no, or very small, effect on the observed radial
distributions as was shown in \citet{tos96}, where a comparison of
models built by different groups was discussed. The only important
characteristic that discriminates among the existing chemical
evolution scenarios is the resulting {\it time evolution} of the
radial gradients of abundances. This time evolution depends on the
ratio SFR/Infall, which becomes the crucial quantity in determining
whether abundance gradients steepen or flatten with time. If infall is
the source of star-forming gas, the gradient flattens as the galaxy
evolves and consumes this gas. Conversely, it steepens with the time
if the effect of the infall gas is mostly that of diluting the
metallicity. In other words, present time H\,{\sc ii} abundances are
not sufficient by themselves to discriminate among the possible models
of disk evolution. Since chemical abundance data other than those
derived from H\,{\sc ii} regions are almost inexistent for external
galaxies, a check on the time evolution of chemical models is only
possible, in principle, for the MWG, where observational data for
individual stars of different ages exist.
This comparison was performed for the MCEM in \citet{mol97}
\footnote{Another possible way to test the time evolution of the models
is of course to calculate the evolution of abundances with redshift
and to compare the predictions with the observations. This was
successfully accomplished for damped Lyman alpha systems
\citep{fer97}, but this procedure does not permit examining the time
evolution of radial {\sl gradients}, or the abundance variations in
different regions of the same disk.}.

An independent test for the evolutionary history of external spiral
disks as predicted by the chemical evolution models was proposed by
\citet[][--hereafter Paper II]{mol99}. There we made use of the radial
distributions of spectral indices Mg$_{2}$ and Fe52, obtained by
\citet{bea97b}, hereafter Paper I, and \citet{bea97} for three
galaxies in the Virgo Cluster, NGC~4303, NGC~4321 and NGC~4535.  The
MCEM applied to these galaxies, reproduced the present radial
distributions and also predicted, for the first time in spiral disks,
the main characteristics of the radial distributions of Mg$_{2}$ and
Fe52 indices. The reader must remember, as pointed out above, that the
calibrated spectral indices represent abundance time-averages over the
galaxy chemical history, as opposed to the accumulated effect shown by
the extreme Pop. I objects such as H\,{\sc ii} regions or the B stars.

But even in the case under discussion we face an uniqueness problem
because the spectral indices synthesis models also show a degeneracy
in the plane Mg$_{2}$-Fe52 (or $\langle {\rm Fe} \rangle$) plane:
Indeed, age produces almost the same effect as metallicity thus
compromising their simultaneous determination for a given single
stellar population, if only these two quantities are used
\citep{bar94,wor94,buz92,buz94}.  Other spectral indices, mostly those
related to the Balmer lines, must be used in order to maximize the
age-metallicity discrimination and to constraint the evolutionary
parameters \citep{fre98,del98}.  The latter studies refer to
elliptical galaxies where a high degree of coevality is
assumed. Spiral disks are subject to a further difficulty in that
stellar disks are composite systems and the time vagaries of star
formation must be taken into account.They are on the other hand richer
in information and physical constraints since data on the gas
abundance and density are available and must be reproduced too.

Thus, there exists a possible ambiguity in that different star
formation histories in disks are possible that would lead to the same
final averaged spectral indices and/or present gas distributions.  Are
there alternatives to the SFR histories found in Paper II which may
reproduce at the same time the present gas radial distributions and
the stellar content information? Do both problems -- uniqueness and
degeneracy -- go in the same or in opposite directions? In other
words, can we constrain the possible sets of evolutionary histories on
the basis of {\it all} the available observations?

In this work, we attack this problem by modifying the input parameters
of the MCEM for the galaxy NGC~4303 for which we compute a large
number (500) of different chemical evolution models.  Then, by using a
$\chi^{2}$ optimization, we choose those able to fit the present-time
nebular observations with probabilities larger than 97.5 \%.  Next,
radial distributions of Mg$_{2}$ and Fe5270 are computed by using the
chemical abundances and star formation histories predicted by the
above--selected evolutionary models as input for the synthesis
procedure employed to calculate spectral indices. We will determine if
the comparison of these spectral indices distributions with the data helps
in the final selection of the evolutionary histories valid over the
disk.

In Section 2 we briefly describe the MCEM.  The computed models as
well as the method of selection among them, based on their predictions
for the present epoch are also in Section 2. The calculation of
synthesis models and the radial distributions of the spectral indices
Mg$_{2}$ and Fe5270, as well as their comparison with observations,
are shown in Section 3. We discuss these results in Section
4. Finally, our conclusions are presented in Section 5.

\section{The Multiphase Chemical Evolution Model (MCEM)}

\subsection{A Brief Description}

We remind the reader that the appropriate check of validity of the
basic multiphase model, here adopted, has already been done in
previous work, leaving open however the issue of uniqueness.
Therefore, we only summarize here the MCEM whose detailed properties,
as applied to spiral disks, are described in \citet{fer92,fer94,mol96}
and in Paper~ II of this series, and concentrate on the comparison
among models.

In the MCEM a protogalaxy is assumed to be a spheroid composed of
primordial gas with total mass M(R). For NGC~4303 the latter is
calculated from the rotation curves obtained either via radio
\citep{guh88} or optical \citep{dis90} observations. The galaxy is
divided into concentric cylindrical regions 1 kpc wide. The model
calculates the time evolution of the halo and disk components
belonging to each cylindrical region. The halo gas falls into the
galactic plane to form the disk, which in the multiphase framework is
a secondary structure formed by the gravitational accumulation of the
diffuse gas, $g_{H}$, at a rate $f$.  The infall rate $f$ is inversely
proportional to the collapse time scale $\tau_{coll}$, which is
assumed to be dependent on galactocentric radius through an
exponential function with a scale length $\lambda_{D}$:

\begin{equation}
\tau_{coll}(\rm R)=\tau_{0}e^{-(\rm R-R_{0})/\lambda_{D}}
\end{equation}

The characteristic time scale $\tau_{0}$ above is defined for every
galaxy as being that of a region located at a radius $\rm R_{0}$,
equivalent to the Solar Neighborhood (hereafter SN) in the MWG, which
is itself scaled from $\rm R_{\rm eff}$ = 62.6 arcsec
(\citet{hen92}). Thus, the adopted radius $\rm R_{0}$ for this galaxy
is 6 kpc. The value for the characteristic time scale may be
determined, using the total mass of the galaxy, through the
expression:

\begin{equation}
\tau_{0} \propto M_{9}^{-1/2} T 
\end{equation}

from \citet{gal84}, where $M_{9}$ is the total mass of the galaxy in
10$^{9}M_{\odot}$ and T is its age, assumed 13 Gyr.  Thus, we 
obtain $\tau_{0}$ for each galaxy, via the ratio:

\begin{equation}
\tau_{0}=\tau_{\odot}(M_{9},gal/M_{9,MWG})^{-1/2}
\end{equation}

where $M_{9,MWG}$ is the total mass of MWG and the value of
$\tau_{\odot}=4 $ Gyr, corresponding to the Solar Neighborhood, was
determined in \citet{fer92}.

In the various regions of the galaxy (halo or disk) we allow for
the different phases of matter aggregation: diffuse gas ({\sl g}), clouds
({\sl c}, only in the disk), low-mass ($s_{1}, m < 4 M_{\odot}$) and
massive stars ($s_{2}, m \ge 4 M_{\odot}$), and remnants. The mass in
the different phases changes through several conversion process, all
of them related to the rate of the corresponding star formation process:

\begin{enumerate}

\item Star formation by the gas spontaneous fragmentation in the halo
($\propto K g_{H}^{1.5}$)

\item Cloud formation in the disk from diffuse gas ($\propto \mu
g^{1.5}$)

\item Star formation in the disk from cloud-cloud collisions 
($\propto H c^{2}$)

\item Induced star formation via massive star-cloud interactions
($\propto a c s_{2}$)

\item Diffuse gas restitution from these processes
\end{enumerate}

The rates for these processes are proportional to the parameters {\sl
K}, $\mu$, {\sl H}, and {\sl a}, which, in turn, depend on
galactocentric radius through the equations derived in FMPD.  The
proportionality factors in these equations are the corresponding
efficiencies of the various processes, that is the efficiency of the halo star
formation, $\epsilon_{K}$, and, in the disk, the probability of cloud
formation, $\epsilon_{\mu}$, of cloud---cloud collision,
$\epsilon_{H}$, and of the interaction of massive stars with clouds,
$\epsilon_{a}$.

The term $\epsilon_{K}$ is assumed constant for all halos, thus being
independent of morphological type. The last term in the list above,
containing the induced star formation, is associated to local
processes and, as a result, its coefficient $\epsilon_{a}$ is
considered independent of both position and morphological
type. However, the other two rates, $\epsilon_{\mu}$ and
$\epsilon_{H}$, depend on the Hubble type and/or the arm class.
Therefore they must be chosen for each galaxy, their variation range
being determined following the arguments given by \citet{fer88}, as
discussed in \citet{mol96} and in Paper II.  These efficiencies
resulted larger for earlier morphological types and lower for the
later ones, and this is useful for the selection of initial values for
a given galaxy as we showed in Paper II. Within this range, a fine
tuning selection for these efficiencies $\epsilon_{\mu}$ and
$\epsilon_{H}$ is performed by choosing those for which the present
time radial distributions (those of oxygen abundances, star formation
rate, and the atomic and molecular gas surface densities) are well
reproduced.  We must stress that although each galaxy has its own set
of efficiencies, the corresponding parameters $\mu$, {\sl H}, and {\sl
K} vary along the galactocentric radius.

The adopted initial mass function (IMF) is taken from \citet{fer90}.
The enriched material is the result of the restitution of processed
material by dying stars and depends on their nucleosynthesis
processes, their IMF (and delayed restitution) and their final fate,
either via quiet evolution or via Type I or II supernova (SN)
explosions.  Nucleosynthesis yields used here are from \citet{ren81},
for low and intermediate mass stars, and from \citet{woo95} for
massive stars, respectively. The type I supernova explosions release
mostly iron, following \citet{nom84} and \citet{bra86}, at a slower
rate than type II, with the result that the iron appears at least 1
Gyr later than the {\sl $\alpha$}--elements (Oxygen, Magnesium ...)
ejected by the massive stars.

The initial conditions are such that when star formation begins early
in the halo zone and later in the disk (a secondary structure in our
model), most of the barionic galactic mass has already accumulated.
Boundary conditions are arbitrarily chosen: the first and simplest
choice is to consider galaxies as closed systems. This choice does not
strongly constraints the internally--driven phenomena -- SN winds for
example are not as prevalent during spiral galaxies evolution as to
produce significant out--gassing, although they can influence internal
redistributions. The environment in which galaxies live may have a
sensibly larger importance \citep{hen93,hen96,pil98}. We discuss here
only {\it closed models}, keeping in mind that this is a first order
approximation.

In the same spirit, we do not include at present in our models the
effect of bars. It is quite evident, even from elementary dynamical
considerations, that barred galaxies must present flatter radial
gradients, as it has indeed been observed
\citep{edm93,mar92,mar94,mar95}.  We will simply consider that an
eventual bar will influence the radial distributions only in the few
internal kpc.

Summarizing, we will retain the same basic scenario applied to other
spirals models but will vary the model input parameters, namely: {\sl the
characteristics collapse time scale, $\tau_{0}$}, according to the
total mass, and the two efficiencies, {\sl $\epsilon_{\mu}$ and
$\epsilon_{H}$}, depending on the morphological type. The combination
of selected parameters will therefore be different for each galaxy.

\subsection{The Computed Models}

Following the arguments above we had estimated in Paper II a value
$\sim 8$ Gyr as the characteristics collapse time scale and a scale
length $\lambda_{D}$ of 4 kpc, for the galaxy NGC~ 4303. After
estimating these parameters defining the collapse law, we chose the
efficiencies $\epsilon_{\mu}$ and $\epsilon_{H}$, (0.22, 0.01), as
those able to reproduce the observed radial distributions for the
atomic gas \citep[H\,{\sc i,}][]{war88,cay90}, the molecular gas
\citep[H$_2$,][] {kenn88,kenn89}, the star formation rate surface
densities obtained from H$\alpha$ fluxes from \citet{ken89}, and
finally the gas oxygen abundances \citep{hen92,mcc85,shi91}. In the
present work will use the oxygen abundances as were corrected by
\citet{pil00}, which seem more adequate.

We proceed now to change the input parameters in order to take into
account the uncertainties in the selection of that first set. Thus,
for instance, the total mass depends on the maximum rotation velocity
which, in some cases is different when calculated either from radio
H{\sc i} or optical data. For example, for NGC~4303, $V_{max}=150$
km/s increases to a value around 216 km/s, from the first to the
second data set. This factor of $\sim 1.5$ in velocity would translate
into a factor of $\sim 2.25$ in $\tau_{0}$. Therefore, we must
increase or decrease the collapse time scale by running models with 5
possible values for the characteristics collapse time scale: $\tau=$
1, 4, 8, 12 and 16 Gyr.

The relation between collapse time scale and mass makes it evident
that $\tau_{coll}$ must change with galactocentric radius. If the
total mass surface density follows an exponential form such as the
surface brightness of spiral disks does, then the resulting
$\tau_{coll}$ must increase along the radius with a scale length
$\lambda_{D} \propto Re$ (where $Re$ is the scale-length for the
surface brightness distribution). This $Re$ varies with the wavelength
band and, moreover, it usually decreases for later types of galaxies
and is larger for the earlier ones. Therefore, although we have so far
used in our standard models $\lambda_{D}=4$ ~kpc for all spiral disks,
we must keep in mind the possibility of variations of this
parameter. Thus, we take other 5 possible values for this scale
length: $\lambda_{D} = $ 1, 4, 8, 12 and 16 kpc.

The effects of changing efficiencies or the collapse time scale are
not equivalent: by increasing the cloud and star formation
efficiencies it is easier to form molecular clouds and also to destroy
them to produce stars, thus resulting in a star formation rate with
higher absolute values, but maintaining the same star formation
history. In other words, the maximum of the star formation rate will
occur always at the same time for models with the same $\tau_{coll}$.
Alternatively, a longer collapse time scale means a lower infall rate
on the disk, resulting into a star formation rate which shifts its
maximum to a later time, producing therefore a slower evolution even
with the same absolute value of its maximum. From Eq.1 the scale
length $\lambda_{D}$, changes the rate of evolution among different
radial regions, increasing the differences among them if it is small
or decreasing them when it is large, thus acting upon the radial
gradients of abundances, which become steeper (or flatter) when the
parameter $\lambda_{D}$ is shorter (or larger).

In order to check the effect of the degeneracy problem in the plane
Mg$_{2}-Fe52$, it is crucial to analyze the differences among the ages
of stellar populations predicted by models with varying collapse times
or length scales, and their effects on spectral indices.

Efficiency values and collapse times are constrained variables that
change together. These probabilistic efficiencies may take values from
0 up 1, and we assume that they depend on the morphological type,
following the equations:

\begin{equation}
\epsilon_{\mu}=e^{-T^{2}/15}
\end{equation}

and

\begin{equation}
\epsilon_{H}=e^{-T^{2}/5}
\end{equation}

where T is the Hubble type \citep[see][]{mol02}, which gives values
similar to those from Paper II where $T=4$, the morphological type
assigned to NGC~4303.

Because these values might change if we vary the infall rate law, we
have also run models with 20 possible values for each set
({$\epsilon_{\mu}$,$\epsilon_{H}$), calculated from the above
expressions, simulating 20 different morphological types from 0 to 10
(by including intermediate values as 0.5, 1.5, etc...). Thus, we ran a
total of 500 models, with all possible variations of these three
parameters, $\tau_{0}$, $\lambda_{D}$ and the set
($\epsilon_{\mu}$,$\epsilon_{H}$) --or equivalently the value of T--.

\subsection{The most probable models from the present day observations}

We must now select which set of parameters among these 500 models are
adequate to represent the observed present-time radial distributions
of abundances, star formation rate, and diffuse and molecular gas
density.

In order to do that, and following the technique used by \citet{tos88}
in her uniqueness problem study, we compute the $\chi^{2}$ indicator,
which measures the proximity of a model to the region limited by
observational error bars:

\begin{equation}
\chi^{2}=\sum_{i=2}^{14}\frac{(Y_{cal,i}-Y_{obs,i})^{2}}{\sigma_{i}^{2}}
\end{equation}

where $Y_{cal,i}$ is the computed quantity, in each radial region $i$,
$Y_{obs,i}$ is the corresponding observed quantity at the same
galactocentric distance, and $\sigma_{i}$ is the error in quantity.

We compute our models for radius from 2 up to 14 kpc in steps of 1
kpc, by avoiding the inner region or bulge because it has a different
geometry and structure.  Therefore, $i$ takes values from 2 to 14.
The value $\sigma_{i}$ is taken as $\sim 0.2$ dex for the oxygen
abundances, the current error in the estimates of abundances, and $\rm
\sim 2 M_{\odot}/pc^{2}$ for the H{\sc i} surface density, a mean
value estimated for these types of data.  For the star formation rate
and the molecular cloud surface density we assume errors as larger as
40\% and 50\%, respectively, due to the larger uncertainties involved
in the derivation via H$\alpha$ and CO fluxes. In particular, we must
take into account that the H$_{2}$ masses in a given region are
estimated through a scaling factor $\alpha$ which depends on the
metallicity of this region \citep{ver95,wil95}. Due to the
uncertainties involved in the calculation of this parameter, the
corresponding radial distribution of $\sigma (\rm H_{2})$ are
considered doubtful \citep{tay01}.

After computing the $\chi^{2}$ indicator for each radial distribution
of oxygen, star formation and gas surface densities produced by our
500 models we must choose the possible models or the range in the input
parameter which may produce models with results falling within the confidence
region. In order to do that, we first select the model corresponding
to the lowest values of $\chi^{2}$, $\chi^{2}_{min}$, for each set of
observations, that is the maximum--probability model for each
constraint.

We now select those models for which $\chi^{2}-\chi^{2}_{min} < \rm
A$, A being the selected value for a given probability which takes
into account the free parameters of the sample and selects the
confidence region for the parameter set From the number of radial
regions, parameters and observational constraints, we search for the
parameter region with $\chi^{2}-\chi^{2}_{min} < 5.62 $, that is
probabilities larger than 97.5\% for each individual data set or
larger than 90\% for all of them over the region of superposition. In
this way, for each model and observational constraint, we have a value
of $\chi^{2}$, and its corresponding probability P.

In order to choose the best models, we must obtain the region of
parameters that minimize their $\chi^{2}$ for each radial distribution
of data separately, and then find the region of coincidence.
Accordingly, we represent these regions of high probabilities for each
observational constraint, all on the same figure so as to delimit the
values of the most probable input parameters. When there are two free
parameters in the theoretical model finding the plane {\sl
parameter1-parameter2} for which each set of constraints has a large
probability of being well fitted, is relatively easy. In practice we
are dealing with a 3 parameters space--$\tau_{0}$, $\lambda_{D}$ and
the set ($\epsilon_{\mu}$ and $\epsilon_{H}$), or its equivalent T
value. Thus, in order to find the region of parameters that reproduce
{\it all} observations with a high probability we will relay on
projections of these planes taking two parameters at a time.

\begin{figure*}
\plotone{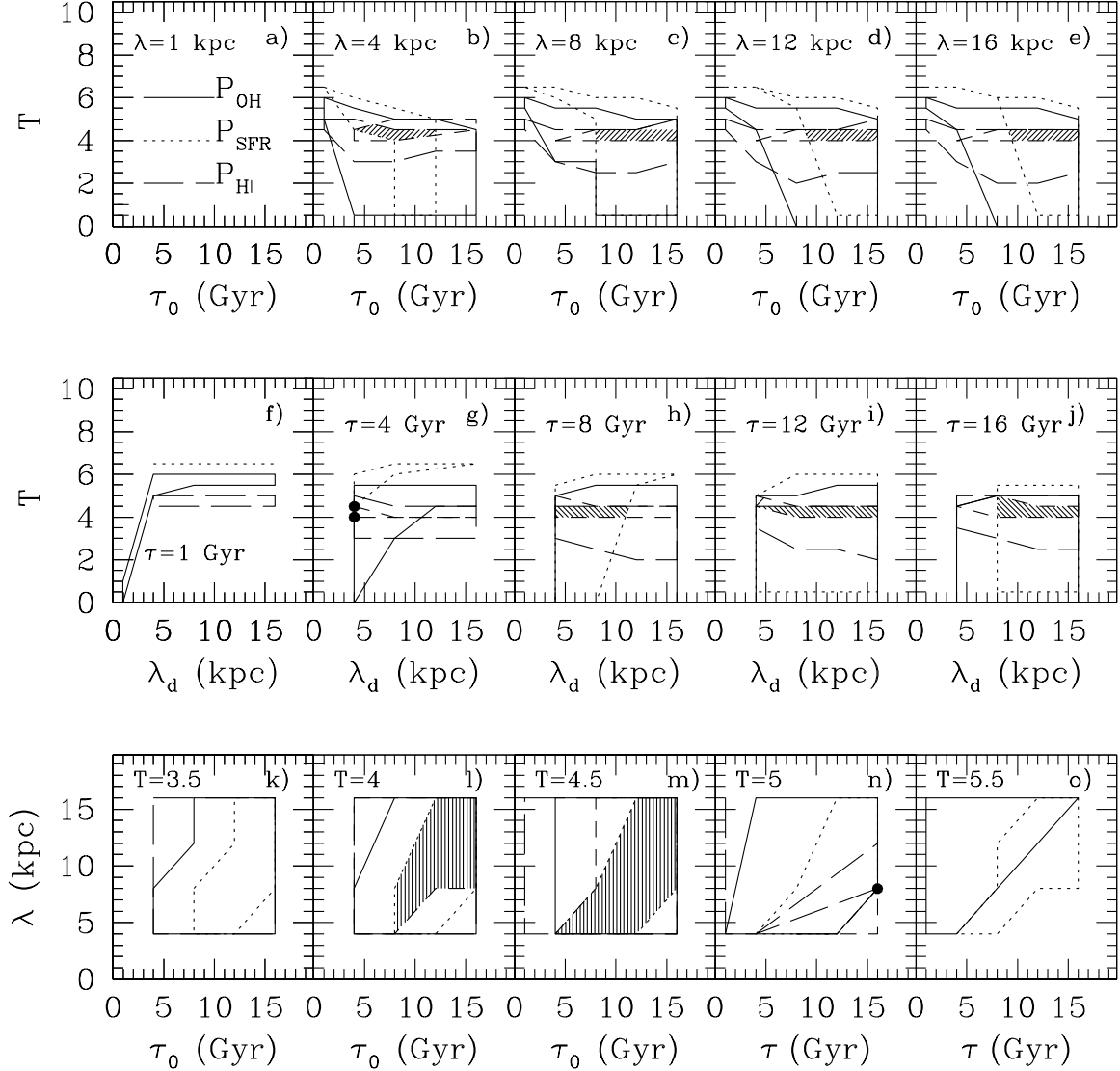}
\caption{Probability contours representing regions where $P > 97.5$\%,
for each observational constraint. Top panels, from a) to e) as a
function of $T$ and $\tau_{0}$. Central panels, from f) to j), as a
function of $T$ and $\lambda_{d}$. Bottom panels, from l) to o), as a
function of $\tau_{0}$, and $\lambda_{d}$.  The solid line represents
the contours for the oxygen abundances. The same for the star
formation rate surface density are the dotted lines, and for the
atomic and molecular gas surface densities are the long-dashed and the
short-dashed lines, respectively.  The shaded regions (or the full
dots when there is no region) are the zones of coincidence of these 4
contours, showing the range of parameters more probable than 90\%.}
\label{prob}
\end{figure*}

In fig.~\ref{prob}, (panels a to e), we show the probability contours
for each radial distribution plotted, as function of type T and of the
collapse time scale $\tau_{0}$ (in Gyr). These 5 panels, one for each
$\lambda_{D}$ value, are the successive projections on the plane
$T-\tau_{0}$ of our 3-dimensional regions along the $\lambda_{D}$
axis. Those parameters which give probabilities larger than 97.5\% for
the oxygen abundance $12 +log (O/H)$ (P$_{OH} > 0.975$) are
represented by solid lines. The dotted lines represent the star
formation rate surface density (P$_{SFR} > 0.975 $) and the atomic and
molecular surface densities ( P$_{HI} > 0.975$ and P$_{H_{2}} >
0.975$), are represented by the long and short dashed lines. The
shaded regions are the coincidence regions where all constraints are
reproduced simultaneously with $P> 90\%$.

In panel a) there are no successful models, all of them produce
probabilities smaller than 97.5\% of fitting the observational
data. It is therefore clear than $\lambda \geq 4$ kpc. The panel with
the largest number of acceptable possibilities correspond to
$\lambda=8$ kpc where the region of coincidence is the largest of the
five. The longer the collapse time scale the larger must be the scale
length. In effect, if $\lambda_{D}= 4$ kpc, $4 < \tau_{0}< 12 $ Gyr;
if $\lambda_{D}= 8$ kpc $ \tau_{0}> 8 $ Gyr, while if $\lambda_{D}=$
12 or 16 kpc, $\tau \ge 10-12$ Gyr. For all these possibilities,
efficiencies fall within a very limited region of $4 \le T \le 5 $.

The latter statement is also apparent from panels f) to j), where the
same kind of probability contours are shown for the plane
$T$-$\lambda_{D}$. From them, we can eliminate the shortest collapse
time scale ($\tau=$1): only if $\tau > 4$ Gyr it is possible to have
coincidence at a level larger than 90 \% for all data. If $\tau= 4 $
Gyr, only a value of $\lambda_{D}=4$ kpc is possible. If $\tau= 8 $
Gyr, then $4 \ge \lambda_{D} \ge 8 $ kpc.  Only for larger collapse
time scales, larger values of $\lambda_{D}$, up to 16 kpc, are also
possible.

>From the graphs above, it is evident that only efficiencies
corresponding to a T between 4 to 5 (i.e., only intermediate and late
Spirals) are possible.  Selection of other efficiency values would
produce poor fittings, mostly in the gas radial distributions, --long
and dashed lines--, for which $\chi^{2}$ increases abruptly if $T < 4$
or $T> 5$.  The bottom panels from k) to o), represent the plane
$\tau_{col}-\lambda$ for T values from 3.5 to 5.5 For values out of
this range, there are again no coincidence regions. It is also clear
from these panels that $\tau_{0} \ge 4-8$ Gyr, and $\lambda_{D}\ge 4 $
kpc are limiting values to obtain good models. Furthermore, it is also
evident that the efficiencies values are strongly constrained by the
gas densities, and that the use of both the diffuse and molecular gas
densities, instead of the total gas mass or the gas fraction, proves
superior in finding models fitting the present-- day data.

The allowed morphological types reduce only to 2 possible values,
$T=4$ or $T=4.5 $, in excellent agreement with the type $T =4$ given
for NGC~4303. If $T=5$ only a model (with $\tau=16$ Gyr and
$\lambda_{D}=8 $ kpc, is able to fit all constraints.  Other T values,
or different efficiencies of diffuse gas and molecular cloud formation
have null probabilities of fitting the gas radial distributions. Since
the morphological type is at least partially defined by the amount of
gas on the disk we do not regard this as a surprising result, but are
reassured that our efficiencies for the different morphological types
are well chosen.

To summarize, the present method yields parameters for which the model
results are within the region defined by the error bars with a
confidence level A. We reduce our possibilities to a $\lambda_{D}$
larger than 4 kpc, a characteristic collapse time scale between 4 and
16 Gyr, and efficiencies corresponding to morphological types T$=$ 4
or 4.5. Only 19 models out of 500 satisfy these conditions, thus
reducing the possible chemical evolution models to a $\sim 4\%$ of the
initial number.

\begin{deluxetable}{lccccccccccccc}
\tabletypesize{\footnotesize}
\tablecaption{Parameters of the Selected Models (P $>$ 90\%) \label{selpro}}
\tablehead{
  \colhead{Model} 
& \colhead{T} & \colhead{$\tau$} 
& \colhead{$\lambda_{D}$} 
& \colhead{$\epsilon_{\mu}$}        &\colhead{$\epsilon_{H}$}
& \colhead{$\chi^{2}_{\rm OH}$}     &\colhead{$\chi^{2}_{\rm SFR}$} 
& \colhead{$\chi^{2}_{\rm H_{I}}$}  &\colhead{$\chi^{2}_{\rm H_{2}}$} 
& \colhead{P$_{\rm OH    }$}        &\colhead{P$_{\rm SFR  }$} 
& \colhead{P$_{\rm H_{I} }$}        &\colhead{P$_{\rm H_{2}}$}}
\startdata
 129&  4. & 4. &   4.5& .259&.017&   0.957&   6.040&   0.378&   2.318&  1.000&  0.983&  1.000&  1.000\\
 228&  8. & 4. &   4.0& .344&.041&   1.201&   2.114&   1.401&   6.413&  1.000&  1.000&  1.000&  0.983\\
 229&  8. & 4. &   4.5& .259&.017&   0.361&   1.716&   1.670&   1.672&  1.000&  1.000&  1.000&  1.000\\
 248&  8. & 8. &   4.0& .344&.041&   3.416&   5.335&   0.265&   4.814&  0.999&  0.992&  1.000&  0.998\\
 249&  8. & 8. &   4.5& .259&.017&   1.922&   5.119&   1.681&   3.177&  1.000&  0.993&  1.000&  1.000\\
 329& 12. & 4. &   4.5& .259&.017&   0.714&   4.887&   2.690&   2.806&  1.000&  0.995&  1.000&  1.000\\
 348& 12. & 8. &   4.0& .344&.041&   2.647&   2.896&   0.689&   5.276&  1.000&  1.000&  1.000&  0.995\\
 349& 12. & 8. &   4.5& .259&.017&   1.223&   2.533&   1.352&   2.101&  1.000&  1.000&  1.000&  1.000\\
 368& 12. &12. &   4.0& .344&.041&   3.398&   4.783&   0.435&   4.985&  0.999&  0.996&  1.000&  0.997\\
 369& 12. &12. &   4.5& .259&.017&   1.937&   4.505&   1.562&   3.238&  1.000&  0.997&  1.000&  1.000\\
 388& 12. &16. &   4.0& .344&.041&   3.773&   6.222&   0.402&   4.947&  0.998&  0.980&  1.000&  0.997\\
 389& 12. &16. &   4.5& .259&.017&   2.336&   6.089&   1.799&   4.032&  1.000&  0.982&  1.000&  0.999\\
 448& 16. & 8. &   4.0& .344&.041&   2.095&   3.567&   1.337&   6.075&  1.000&  0.999&  1.000&  0.988\\
 449& 16. & 8. &   4.5& .259&.017&   0.858&   3.219&   1.130&   2.056&  1.000&  1.000&  1.000&  1.000\\
 450& 16. & 8. &   5.0& .189&.007&   1.506&   2.976&   5.057&   3.874&  1.000&  1.000&  0.989&  1.000\\
 468& 16. &12. &   4.0& .344&.041&   2.841&   3.930&   0.804&   5.543&  1.000&  0.999&  1.000&  0.993\\
 469& 16. &12. &   4.5& .259&.017&   1.440&   3.480&   0.918&   2.611&  1.000&  1.000&  1.000&  1.000\\
 488& 16. &16. &   4.0& .344&.041&   3.215&   4.675&   0.634&   5.387&  0.999&  0.996&  1.000&  0.994\\
 489& 16. &16. &   4.5& .259&.017&   1.795&   4.237&   0.965&   3.167&  1.000&  0.998&  1.000&  1.000\\
\enddata
\end{deluxetable}

We show in Table~\ref{selpro} $\chi^{2}$ and the corresponding
probability P for each one of the observational constraints, for these
19 models which have probabilities larger than 97.5\% to be close to
the observations (which in turn corresponds to 90\% for the
combination).  There we have in column (1) the number identifying our
model. Column (2) is the value of T. The characteristics collapse time
scale, $\tau_{0}$, and the scale length, $\lambda_{D}$, are in columns
(3), and (4). Efficiencies $\epsilon_{\mu}$ and $\epsilon_{H}$ are
given in columns (5) and (6).  In Columns (7) to (10) the values of
$\chi^{2}$, corresponding to the radial distributions of abundances,
star formation rate, and diffuse and molecular gas surface densities,
are given. Finally, the probabilities of these distributions to be in
a region around the minimum value of $\chi^{2}$ are in columns (11) to
(14).  We conclude there is only a model with $\tau_{0}=4$ Gyr and
only a model with $T=5$. All the others correspond to $T=4$ or 4.5
with $\tau_{0}\ge 8$ Gyr.

\subsection{Present Day Radial Distributions}

Following the procedure described above we have found 19 different
models for N4303 satisfying the condition that the radial
distributions of abundances, star formation rate and gas densities all
have probabilities larger than 90\% of fitting the observational data.

\begin{figure*}
\plotone{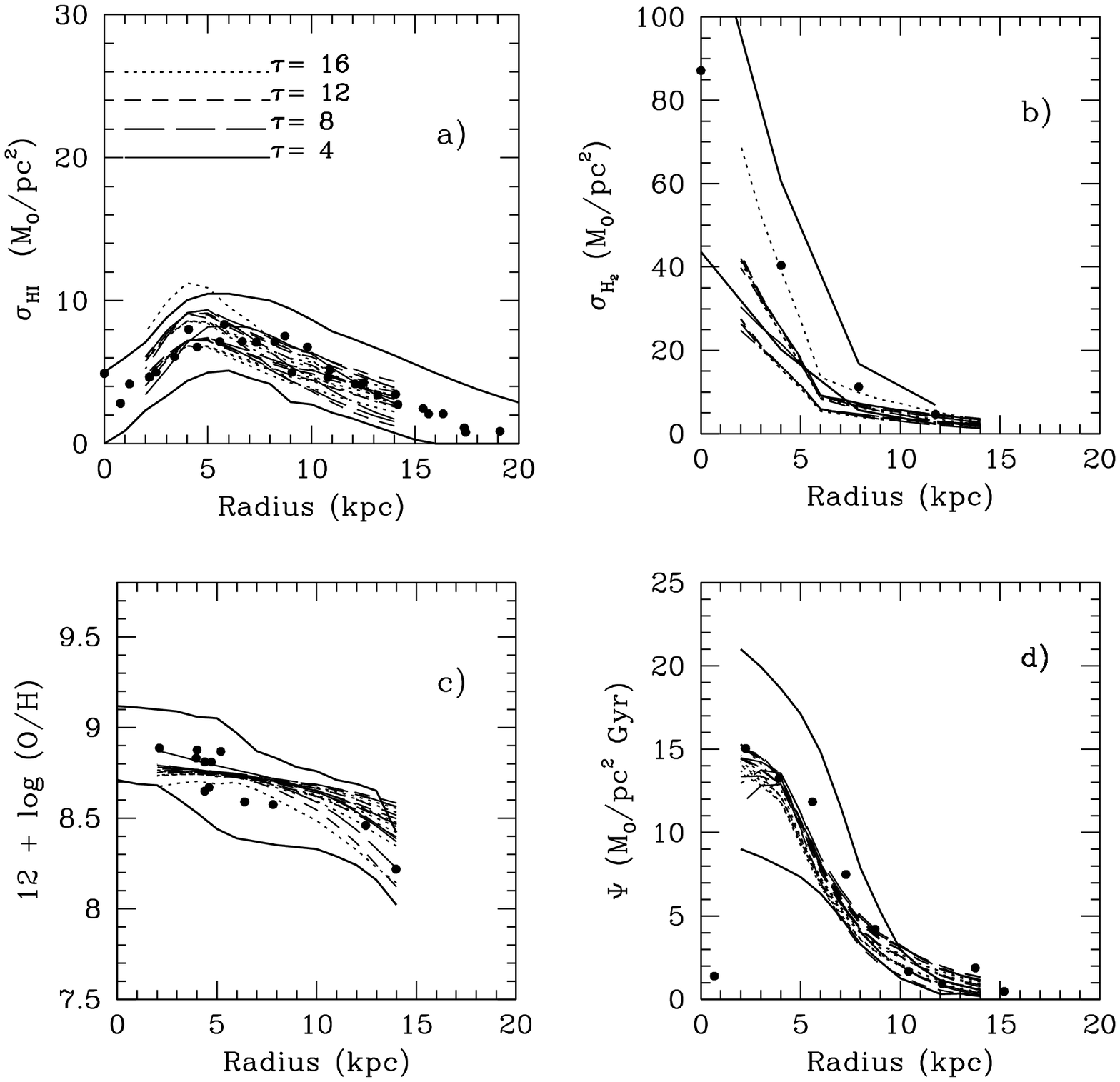}
\caption{Predicted Radial distributions for models which satisfies the
condition P$> 90 \%$ simultaneously for all present time data: a)
diffuse or atomic gas H\,{\sc i} surface density; b) molecular gas
H$_{2}$ surface density; c) Oxygen abundance as $12+log (O/H)$ and d)
Star Formation Rate surface density. Models are represented by
different lines and symbols following the figure. Observational data
are the full dots while the thick lines limit the region within
observational error bars used in our realizations.}
\label{selt}
\end{figure*}

We represent all models in Fig.~\ref{selt} where we see that all of
them are falling within the regions defined by the observational
limits, obtained with the data and the error bars and represented by
the thick solid lines.  We regard the latter subset of models as
equally probable within the errors and seek now additional constraints
to isolate the best models among them, representing the true
evolutionary history of the galaxy.

\section{Using Evolutionary Synthesis Models}

Unlike H{\sc ii} abundances, whose abundances reproduce the cumulative
effect of all past generations of stars, stellar indices contain
information on the evolutionary history of a galaxy in the form of an
average over time and over the SFR. They provide therefore additional
constrains to our models and we show in this section how to retrieve
this information.

\subsection{The method of calculation}

As pointed out, for N4303 (and two other galaxies of paper II) in
addition to the typical quantities necessary to define the galactic
evolution (i.e., gas, metals,...), we have at our disposal
measurements of radial distributions of the spectral indices Mg$_{2}$
and Fe52.  The fact that the models selected in the preceding
sections represent well the present-day radial distributions, does not
imply that the evolutionary history given by the multiphase model are
unique, as no time-dependent information has been used so far. In
fact, we have obtained results in agreement with the present gas data
either with fast evolution (such as Model 129 with $\tau_{0}=$ 4 Gyr)
thus producing stellar populations biased toward older ages and an
earlier enrichment, or with a slower creation of stars (such as Model
450 $\tau_{0}=$16 Gyr) thus producing metals at a more constant
rate. Both alternatives yield almost the same present radial
distributions.  Would they also produce similar radial distribution of
stellar spectral indices?  We will now compute these indices for each
of our models in an attempt to answer this question, and will try to
check if the radial distribution of spectral indices are equally well
reproduced by the adopted set of models.

We proceed in a self-consistent way as follows. Using the chemical
evolution models already described we calculate the integrated mass of
stars created in a given time interval by the gas, and the mean
abundance reached at that epoch by the same gas. We consider the
resulting stellar populations residing in every galactocentric region
as the superposition of a set of single stellar populations or {\sl
generations}, each one of which is defined by its age and its
metallicity. We remind the reader that our models provide abundances
for 15 elements, including Oxygen, Magnesium and Iron, so we know in
addition the abundance ratio [Mg/Fe] for each generation. We have in
principle the necessary ingredients to evaluate observable spectral
indices {\sl via} a synthesis model.

Spectral index features are calculated with the same method described
in \citep{mol00b} by using the Padova group isochrones. The technique
in question was developed to be used specifically in spiral disks or
in regions where a combination of stellar populations of different
ages and/or metallicities is necessary, and it works by simultaneously
computing the spectral features and the continuum flux. An isochrone
is assigned to each of our stellar generations using the total
abundance Z reached at this time step and the corresponding age. Then,
an index value is calculated for each star belonging to that
generation following its stellar gravity $\log g$, effective
temperature $\rm T_{eff}$ and metallicity or abundance, through the
fitting functions, and a synthetic spectra from atmosphere models is
selected. By adding all stellar contributions, the contribution of
every generation to both synthetic single continuum and line fluxes is
computed at the same time. Next, continuum and line fluxes are added
for all generations, weighted by the stellar mass created in every
time step, according to the star formation history provided by the
evolution model. This calculation is performed for all disk regions,
generating spectral indices along the galaxy disk.

We use the fitting function from \citet{wor94b} for assign the index
Fe52 to each star. However, the fitting function for $Mg_{2}$ must
depend on the abundance [Mg/H], not [Fe/H], contrary to what is
assumed in that work. The selection of [Mg/H] or [Fe/H] in spiral
disks is important due the different rate in the ejection of these
elements and the differences among the disk regions along the radius.
Alternatively, \citet{bor95} obtained a fitting function for the index
$Mg_{2}$ by including a dependence on [Mg/H]. However this dependence
is derived from the relative abundance [Mg/Fe].  We prefer to use
directly the dependence on [Mg/H]. Therefore, by using the same the
stellar library from these authors, we have derived our own fitting
function which gives us the direct dependence of Mg$_{2}$ on [Mg/H].
The abundance of [Mg/H] is used to calculate Mg$_{2}$ and the
abundance of [Fe/H] to calculate Fe52.

\subsection{Spectral indices results}\label{indres}

We have computed in this way the radial distributions of spectral
indices for all models of the galaxy NGC~4303 with larger than 90\%
probability in all observational constraints for the present-time
data, as displayed in Table~\ref{selpro}.

Can the comparison with spectral index radial distributions help us
further in selecting the best evolutionary model or models, or limit
the range of parameters of these?.

\begin{deluxetable}{rccccccc}
\tabletypesize{\footnotesize}
\tablecaption{Results from synthesis models.\label{chiind}}
\tablewidth{0pt}
\tablehead{
\colhead{Model}  &       
\colhead{$\chi^{2}_{\rm Mg_{2}}$} & \colhead{$\chi^{2}_{\rm Fe52}$} &   
\colhead{P$_{\rm Mg_{2}}$} & \colhead{P$_{\rm Fe52}$}&
\multicolumn{3}{c}{Radial Gradients}\\
\colhead{Number} &\colhead{} & \colhead{}& \colhead{} &\colhead{} & 
\colhead{$\bigtriangledown_{\rm O/H}$} & 
\colhead{$\bigtriangledown_{\rm Mg_{2}}$} &
\colhead{$\bigtriangledown_{\rm Fe52}$}
        }
\startdata
$\ast$ 129 &  5.30 &   3.17 & 1.00&  1.00 & -0.079& -0.0062& -0.094 \\
$\ast$ 228 &  6.69 &   3.94 & 1.00&  1.00 & -0.076& -0.0048& -0.074\\
$\ast$ 229 &  6.60 &   4.89 & 1.00&  1.00 & -0.086& -0.0050& -0.078\\
       248 & 11.06 &   3.49 & 0.97&  1.00 & -0.050& -0.0030& -0.045\\
$\ast$ 249 & 10.51 &   4.41 & 0.98&  1.00 & -0.059& -0.0037& -0.056\\
$\ast$ 329 &  8.30 &   6.75 & 1.00&  1.00 & -0.091& -0.0048& -0.075\\
       348 & 11.80 &   5.02 & 0.95&  1.00 & -0.055& -0.0031& -0.047\\
$\ast$ 349 & 10.70 &   6.29 & 0.98&  1.00 & -0.063& -0.0035& -0.054\\
       368 & 12.81 &   4.90 & 0.91&  1.00 & -0.047& -0.0023& -0.034\\
       369 & 13.41 &   4.82 & 0.88&  1.00 & -0.054& -0.0021& -0.031\\
       388 & 13.41 &   4.82 & 0.88&  1.00 & -0.043& -0.0021& -0.031\\
       389 & 13.17 &   6.01 & 0.90&  1.00 & -0.050& -0.0028& -0.042\\
       448 & 12.19 &   6.04 & 0.94&  1.00 & -0.058& -0.0030& -0.046\\
       449 & 14.38 &  10.92 & 0.83&  0.90 & -0.066& -0.0037& -0.057\\
       450 & 14.38 &  10.92 & 0.83&  0.90 & -0.072& -0.0037& -0.057\\
       468 & 13.13 &   5.94 & 0.90&  1.00 & -0.050& -0.0025& -0.037\\
       469 & 12.49 &   7.41 & 0.93&  0.99 & -0.057& -0.0030& -0.047\\
       488 & 13.61 &   5.87 & 0.87&  1.00 & -0.046& -0.0023& -0.034\\
       489 & 12.73 &   7.32 & 0.92&  0.99 & -0.053& -0.0029& -0.043 \\
        OBS&       &        &     &       & -0.078& -0.0086 & -0.073\\
\enddata
\end{deluxetable}

We compute once again the statistical indicator $\chi^{2}$ but this
time for each one of the two distributions as given by spectral
indices.  The results of these calculations are shown in
Table~\ref{chiind}, where we give the goodness of fit for each
model. For each model number given in Column (1), Column (2) gives
$\chi^{2}$ for the Mg$_{2}$ data, Column (3) for Fe52. The
corresponding probabilities are in Columns (4) and (5). The radial
gradients obtained for each model and by the observational data
distributions are also given in Columns (6),(7) and (8).

\begin{figure*}
\plotone{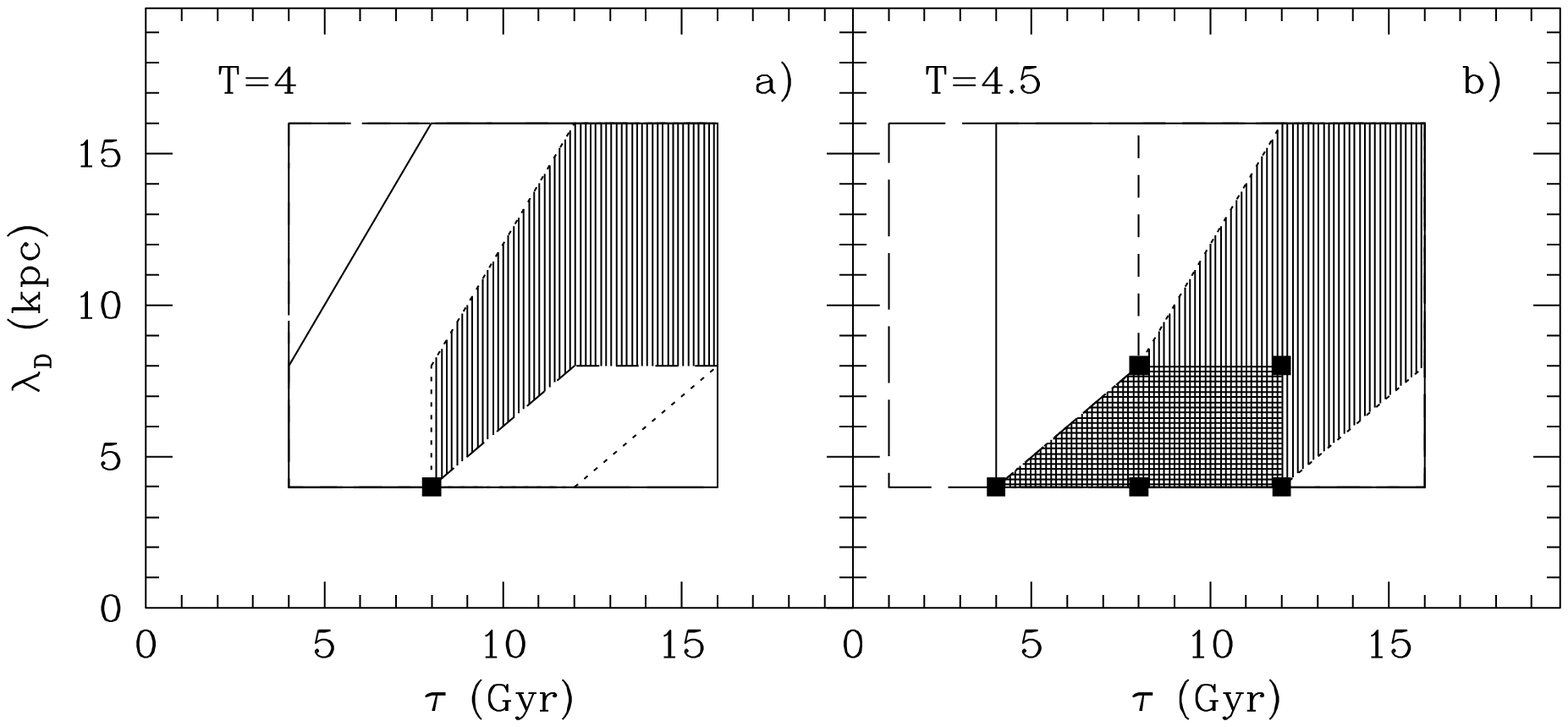}
\caption{Probability contours representing regions with P$ \ge 97.5$ \%
 for all observational constraints for T$=4$ (panel a) and T$=4.5$
(panel b) as a function of $\tau$ and $\lambda$. Lines represent the
same than panels l) and n) from Fig.~\ref{prob}, while filled squares
 are the models obtained in Section~\ref{indres}}
\label{taul}
\end{figure*}

In Fig.\ref{taul} we represent the same panels l) and m) from
Fig.~\ref{prob} but limiting those resulting regions with these new
constraints. In panel a), T$=4$, we see that there is only a point,
this one corresponding to Model 129, which satisfies our confidence
condition of 97.5\% for all observational constraints. For panel b),
T$=4.5$, there still exists a little region shaded by both vertical
and horizontal lines (that one shaded only by vertical lines
correspond to the region obtained with chemical evolution models)
showing the possible input parameters which reproduce all data with
the same goodness. There are more than one model, but we have restricted
the possibilities to a very reduced zone.

The best model fitting the observational data with $P=1$ for all 
constraints , corresponds to Model 229, with a
morphological type $\rm T =4.5$, a collapse time scale $\rm
\tau_{coll}=8$ Gyr and $\lambda_{D}=4$ kpc, a model similar to Model~
B presented in Paper II as the best model. Only 5 other models, marked
with an $\ast$, have also able to fit these new constraints with
probabilities larger than 97.5\%.  In particular, Model 129, the only
one among the 19 with $\tau_{0}=4$ kpc, have also high values of
probabilities.  All the other possible models, mostly in the low part
of the table, that is with slow collapses and/or low star formation
efficiencies, although reproduce the present day radial distributions,
are far away from our confidence region for one or both spectral
indices. Thus, Model 450, with P$\ge 0.99$ in Table~\ref{selpro}
decreases its probabilities, when the comparison of model predictions
with data is performed through the spectral indices Mg$_{2}$ and Fe52,
to 0.83 and 0.90, respectively.  Therefore, we reduce to 6 the
possible models from the initial 500, this is only 1 \% of them are
retained by the procedure.

We may answer to the stated question, can the comparison with
spectral index radial distributions help us to select the appropriate
evolutionary model? with an affirmative statement.  There are some
sets of physically plausible input parameters that provide an adequate
enough fit to the abundance data.  However, these models can not
reproduce equally well the spectral indices radial distributions.

\begin{figure*}
\plotone{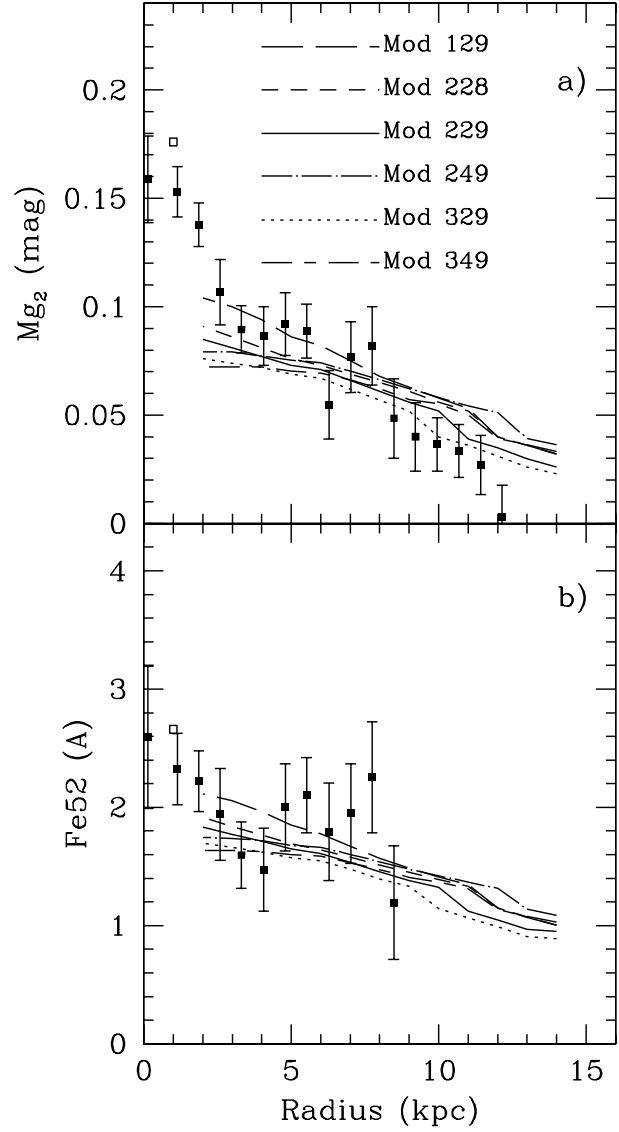}
\caption{Spectral indices radial distributions for NGC~4303. Panel a)
the index Mg$_{2}$ in magnitudes. Panel b) The index Fe$_{52}$ in
\AA.  Data are from \citet{bea97} and \citet{bea97b}.}
\label{indices}
\end{figure*}

We show in Fig.~\ref{indices} the predicted radial distributions of
Mg$_{2}$ (in panel a), and Fe52 (in panel b), respectively, for these
6 models.  In the graphs of Fe52, observational points in the outer
disks (R $>$ 10 kpc) of all galaxies tend to increase with radius; we
think that some of these points at large and moderate radii may be
affected by large systematic uncertainties in the sky-subtraction
procedure. The comparison of models with these Fe52 outer data is not
significant.

For both indices there is a inner region where the observed values
raise significantly. This region seems to contain a stellar bar which
cannot be simulated with our present models, and as discussed, we have
not included these regions in our calculations.  However, we have 
added the results obtained for the bulge or central region with the
Model B from Paper II in \citet{mol00}.  The only difference between
this bulge model and the disk models used here resides in the geometry
and structure of this region.  The input parameter for this bulge
model are very similar to Model  229. Taking into account
that the most of differences among models appears in the the outer
disk predictions, we can assume that by using the 5 other model input
parameters in the bulge model would give similar results for this
bulge or central region. These results are shown in Fig.\ref{indices}
as open squares located in a mean radius for the bulge, assumed as 1
kpc. We have not computed the complete model with the full radial
range included, for which a model allowing the calculation of radial
gradients in the bulge is necessary.  We believe that these central
values fall on the right absolute level, and that the shape of the
spectral indices distributions in the inner disk may be also well
reproduced.

Summarizing, the uniqueness issue associated the chemical
evolution models is not strong, because the models were already
reduced to a 5\% with just the present day data as constraints. But
when we use in addition the spectrophotometric indices we limit even more
the possible evolutionary histories.

\subsection{The origin of the radial variations of Magnesium and Iron}

What is the explanation for the strong observed Mg2/Fe52 radial
gradient?  Indeed, computing the Mg2/Fe52 ratio from the data in
columns 6,7 and 8 of Table~\ref{chiind} the gradient turns much more
steeper in the observational data than predicted by the models. We
have no straightforward explanation for this behavior.

From a chemical evolution point of view Magnesium appears early in the
evolution, as it is produced by massive stars which evolve
rapidly. Iron, instead, is created by the explosions of SN-I which
need a time delay to appear. Although abundances tend to a saturation
level, saturation is reached earlier for Magnesium, and as a result
the Iron radial gradient subsists longer than the Magnesium one.

When stellar averaged abundances are computed, taking into account
that star formation rate proceeds more rapidly in the inner disk than
in the outer regions, an abundance flattening results which is very
similar for both indices.  In fact, the averaged stellar relative
abundance $\rm <[Mg/Fe]>$ is almost zero for our 6 models, without
appreciable variations along the disk.

We must remember, however, that spectral indices are not directly
equivalent to abundances because they also depend on the mean age of
the parent stellar populations. Therefore, one might think that the
explanation for the strong radial gradient of the ratio Mg2/Fe52 could
reside in the different dependence on age for both indices. It is well
known in fact that Mg2 changes with age more strongly than Fe52.  This
dependence is however already included in the synthesis models through
the fitting functions which include a component that varies with the
effective temperature of stars. Thus, the models should be able to
predict the observational data.

Maybe our models do not contain a sufficient variation in the mean age
of the stellar populations along the radius. It is possible to obtain
stronger radial gradients in age by decreasing the collapse time
scale, for example $\tau_{coll}=1$ Gyr, or by taking a shorter scale
length $\lambda_{D}$.  Both possibilities have been already included
in our initial 500 models. But it should be noted that these types of
models evolve very rapidly in the inner disk by swiftly consuming all
the gas, even the molecular phase. The resulting radial distributions
of molecular gas density show a hole in the center in clear
disagreement with the observed exponential shape. It is also useful
to remember that the uncertainties in the estimations of the molecular
phase mass is large, due mostly to the dependence of its calibration
on the metallicity of the region under study. We suspect that if this
dependence is included, the density of molecular gas in the inner disk
could decrease and the models would change accordingly.

Other possible solution is that the age dependence of Mg2 or Fe52
included in the fitting functions is inadequate. The fitting function
for Mg2 from \citet{bor95} has a stronger dependence on age than the
one from \citet{wor94}, but the parameters range for the first
library is narrower that for the second one. The latter may produce
problems when young populations coexists with the older ones, as is
the case in spiral disks. This is particularly so in the outer
regions, precisely where models differ more from the observations.

In our simple scenario the disk is created from the collapse of the
gas of the protogalaxy.  One might however think that other mechanism
such as mergers are also able to produce variations in the ratio of
Magnesium to Iron.  Existing models for elliptical galaxies seem to
indicate that in the merger scenario a ratio $\rm [Mg/Fe]>0$, and larger
in the center, cannot be obtained \citep{tho00}. In fact the observed
trends in [Mg/Fe] are the opposite to those derived from the
semianalytical models simulating mergers of spiral to form
ellipticals.

Finally we must take into account the uncertainties associated with
the difficult measurements of external disk regions of low surface
brightness.  It is clearly necessary to obtain a larger dataset of
absorption spectral indices in spiral disks for galaxies with strong
radial gradients of nebular abundances.

\section{Discussion: the possible evolutionary histories}

The impact of radial distributions of spectral indices on the model
predictions can be explained through the effect of the star formation
histories on the average stellar abundances. The star formation
histories of our 19 models are different and this must imply different
mean ages for the stellar populations along galactocentric radius.

\begin{figure*}
\plotone{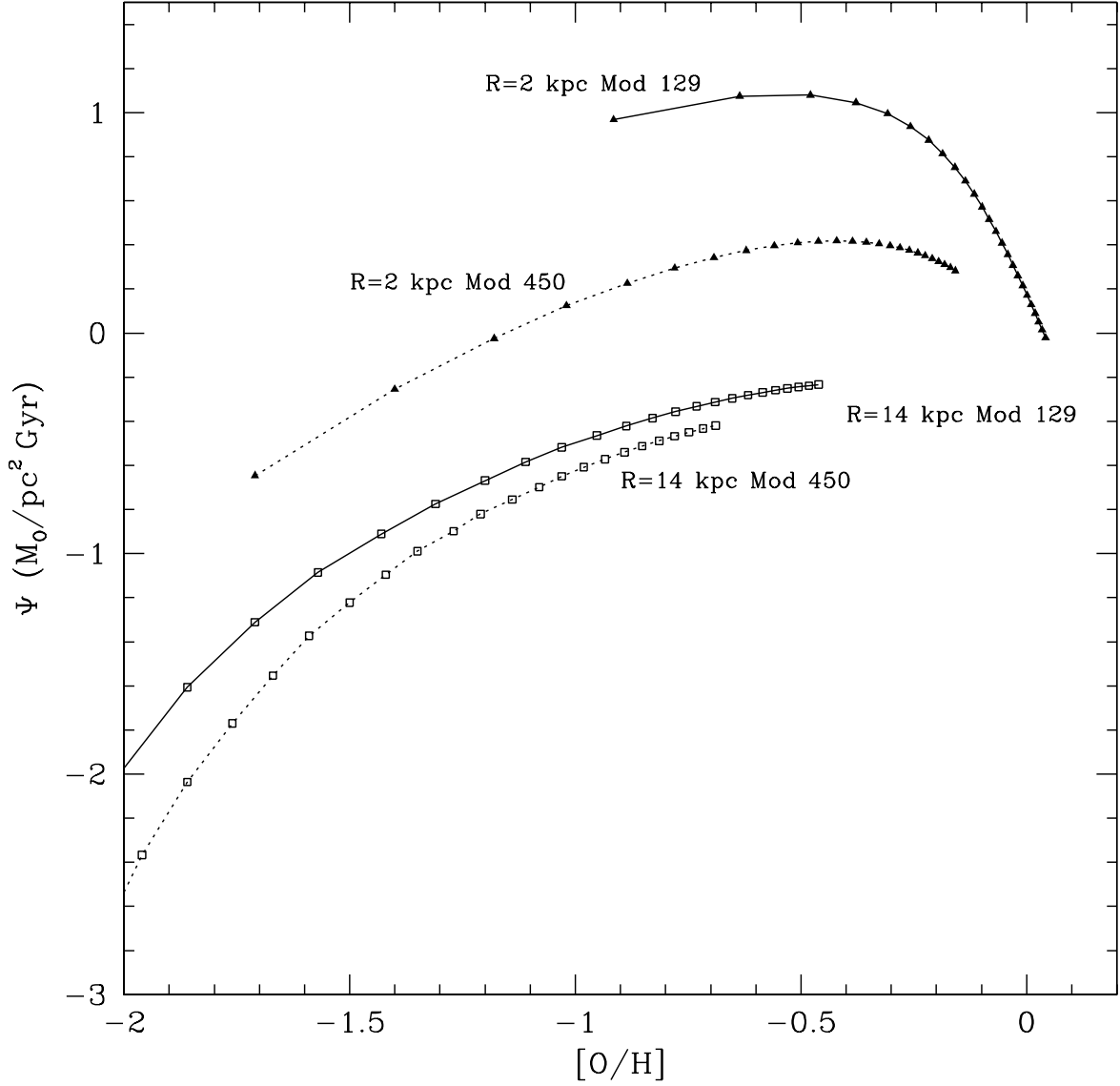}
\caption{Evolution of the star formation rate $\Psi$ {\sl vs} the
oxygen abundance for the inner ($R=2$ kpc, $\triangle$ symbol) 
and outer ($R=14$ kpc, $\square$ symbol) regions  for Models 129 (solid lines)
 and 450 (dotted lines).}
\label{sfr-z}
\end{figure*}

This may be clearly seen in Fig.~\ref{sfr-z} where we represent the
star formation history for a inner ($R=2 $ kpc) and a outer ($R=14 $
kpc) disk regions {\sl versus} the oxygen abundance for two models. We
have selected for comparison Models 129 and 450, which we consider
clearly separated in the input parameter space. The first one is
represented by the solid line while the dotted line shows results for
the second one. The inner regions have triangles as symbols, and the
outer squares. We see that Model 129 is above Model 450 for both
regions. Although the final results are not very different, the
evolutionary histories truly are, mostly in the inner
regions. Therefore mean ages and abundances produced by these models
must differ, and, as a result, the spectral indices must be different.

We have computed these averaged ages and abundances for these two
regions of the 19 models from Table~\ref{selpro}. We represent our
results in Fig.~\ref{tau_z}, panel a) for $R=2$ kpc and panel b) for
$R=14$ kpc.

\begin{figure*}
\plotone{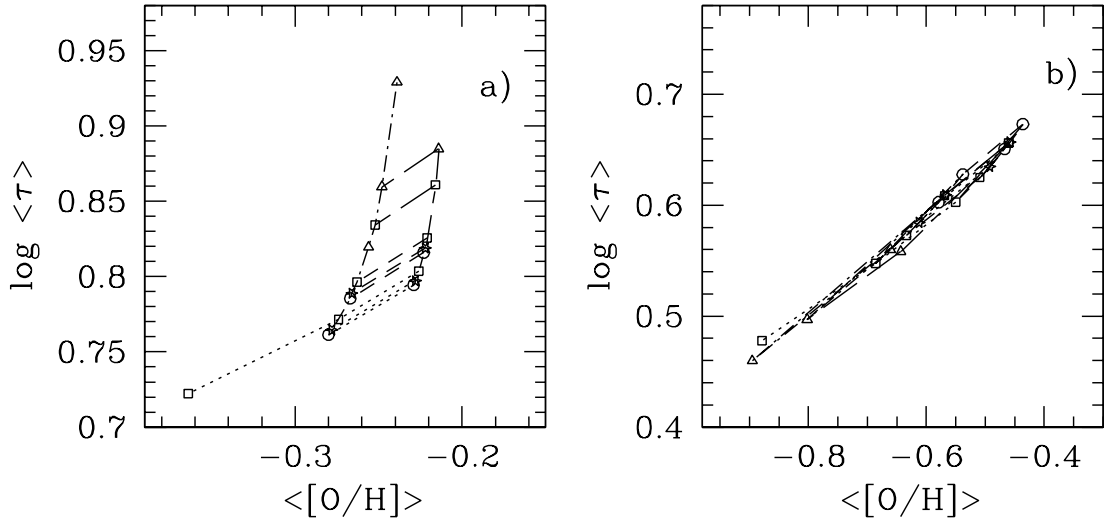}
\caption{The corresponding averaged age of the stellar populations
{\sl vs} the averaged oxygen abundance $<[O/H]>$ for the 19 models
in 2 radial regions: a) R=2 kpc and b) R= 14 kpc.  The two
vertical lines in panel a) join models with $T=4$ (dot-long dashed line) and
$T=4.5$ (short-long dashed line). The square at the left , Model 450,
is $T=5$. Horizontal lines represent $\tau_{0}=8$ (long-dashed line), 12
(short-dashed) and 16 Gyr (dotted line), respectively. The top square
is $\tau_{0}=4$ Gyr.  Symbols mark different $=\lambda_{d}$:
$\triangle$ for $\lambda_{d}=4$; $\square$ for $\lambda_{d}=8$;
$\bullet$ for $\lambda_{d}=12$ and $\star$ for $\lambda_{d}=16$ kpc,
respectively. Panel b) Symbols have the same meaning than in Panel a).}
\label{tau_z}
\end{figure*}

In panel a), with a point for each model, we see an almost
orthogonal diagram.  The near--horizontal lines join models with a same
collapse time scale, with different symbol following the scale length
value ($\triangle$ for $\lambda_{D}=4$; $\square$ for $\lambda_{D}=8$;
$\bullet$ for $\lambda_{D}=12$ and $\star$ for $\lambda_{D}=15$),
while the vertical lines correspond to different efficiencies or T's.
Spectral indices depend on both the abundances and the mean ages of the stars
which produce them, and we see that Model 129, located at the top of
the panel as a triangle, is older and more metal-rich that Model 450,
the square located at the left of the same panel. The corresponding
spectral indices can not be equal. The other models are located in
intermediate positions.  The situation for outer regions ( $R=14$ kpc)
is shown in panel b) where all models fall in a line.

We can see, however, that for both panels the larger the mean
abundances the older the stellar population.  This means that the
degeneracy problem of Mg$_{2}$-Fe52 is not at work here because the
result goes in the opposite direction.  If the abundances turned out
lower in a model than in other we would need a mean age older for the
first in order to obtain similar spectral indices, while in fact our
low-metal rich models turned out to be the youngest ones.  Spectral
indices, specially the Mg$_{2}$ index, which has a strong dependence
on mean age and has a better measurement, will probably discriminate
among models better than the gas abundances. This effect is stronger
in the outer regions of the disk, as we show in panel b), implying
that spectral indices radial distributions may be similar in the inner
regions but not in the external ones, where the gas distributions may
diverge. This explains the effect shown in Fig.~\ref{indices}.

The star formation histories have two effects on spiral disks: a
radial gradient of abundances and a variable mean age for the stellar
populations created in each radial region. Thus, even if the radial
gradient of abundances obtained from different models are close enough
to prevent adequate discrimination, a radial gradient of ages may
still exist.  Since spectral indices have a dependence on age and on
abundances, a radial gradient in the mean ages of the stellar
populations implies that radial distributions of indices for different
models must differ more than those of abundances, thus helping in
selecting the most adequate model or, at least, determining if some of
these fall out of the confidence region of the plane defined by the
data and their error bars.

\section{Conclusions}

We have discussed some chemical evolution models and their comparison
with data. A first type of models, based purely on the multiphase 
chemical evolution models that
yields direct nebular abundances were constrained by the radial
distributions of both atomic and molecular gas.  These observational
data for the present time (Fig.~\ref{selt}), may be reproduced by only
19 among 500 initial models.  Put differently, there are some restricted sets of
physically plausible input parameters that provide a superior fit to the
abundance data.

Next we computed the spectral index values predicted by the same
models through the use of evolutionary synthesis models.  These
computations relay on a synthesis procedure capable of establishing the
abundance-index relation and which is implemented via the application
of the first category of models above. This second category of models, which
depends on the first one, may further constrain the chemical evolution
of spiral disks by introducing measurements of their stellar content
via the determination of spectral absorption indices, such as the
Mg$_{2}$ and Fe5270 Lick indices. Once again we show that the
multiphase model reproduce the generic characteristics of both types
of observed radial distributions. The 19 models initially
selected on the basis of the nebular data do not reproduced equally
well the radial distributions of spectral indices, as can be seen in
Table~\ref{chiind}: only 5 of them are now good enough to satisfy our
criteria.

 We also show that statistical differences among models increase when
spectral indices radial distributions are used with respect to those
using only the present--day oxygen abundances. This effect appears
because, even if the present abundances are similar for different
models, the averages abundances and ages of the stellar populations
responsible for the corresponding spectral indices are not.  In fact,
in a diagram {\sl abundance-age}, our results go in the opposite
direction than expected from the age-degeneracy problem.

Therefore, by taking into account the results of Table~\ref{chiind},
we demonstrate that spectral indices used as time constraints on the
evolution of the galaxy are strong enough to eliminate a significant
fraction of the models, or to help in the selection of possible
scenarios, and we suggest that the measurement of spectral indices
must be considered of importance in designing observational projects
 dealing with the structure of spiral disks.

\begin{acknowledgements}
We thank Dr. A.I. D\'{\i}az for useful suggestions.
This work has been partially supported by DGICYT project AYA-2000-093.
M.Moll\'{a} acknowledges the Spanish {\sl Minister\'{\i}o de
Educaci\'{o}n y Cultura} for its support through a post-doctoral
contract.  This work has been use of the Nasa Astrophysics Data
System, and the NASA/IPAC Extragalactic Database(NED), which is
operated by the Jet Propulsion Laboratory, Caltech, under contract
with the National Aeronautics and Space Administration.

\end{acknowledgements}

\end{document}